# Using the Multirhodotron as an Advanced Rhodotron

M.V. Gavich      Canada      (mgavich@gmail.com)
V.T. Gavich      Montenegro      (vgavich@gmail.com)

**Abstract**

This article assesses the use of a new type of electron accelerator, the Multirhodotron, for four new purposes that cannot be implemented using Rhodotrons and linacs. This study awards some arguments about possible replacement of nuclear reactors by electron accelerators in process of producing of medical isotopes on a global scale, about new possible electron accelerator for high energy, also describes new possibilities of actuation of free-electron lasers (FEL) at the megawatt level in the continuous wave (CW) mode, and suggests some decision of the use of Multirhodotron for technology of "electron cooling" in proton accelerators and colliders.

Key words: Multirhodotron; Rhodotron; electron accelerator; powerful FEL; medical isotopes, "electron cooling"

## Introduction

More than 30 years ago, Pottier and Nguyen suggested a new idea for the acceleration of an electron beam in the radial electrical field of a coaxial resonant cavity energized by the TEM1 mode [1]. Accelerators of this type were called Rhodotrons. These accelerators have the highest power and electrical efficiency among all electron accelerators. Rhodotrons have a high exit power with 60-70% efficiency because unlike the linac, the electron beam in the Rhodotron crosses the same cavity several times but the losses in the cavity walls are considered only once. Increasing the number of electron beam passes through the cavity will further increase the efficiency.

## Rhodotron's and Multirhodotron's features

Rhodotrons allow only 7-12 passes of an electron beam through the cavity because all of the electron beam trajectories occur only in the cavity's single middle plane. This vulnerability to Pottier and Nguyen's patent can be easily eliminated if the electron trajectories are placed in several planes located perpendicular to the axis of the coaxial cavity. We suggested this in 2012 in our patent (CA 2787794).The length of the coaxial cavity can be increased to $\lambda$, 1.5 $\lambda$, 2 $\lambda$, and further, and the cavity can be energized in higher modes such as $TEM_n$, n = 1, 2, and 3. However, in all of the cavity variant planes, the distribution of the electromagnetic fields will be the same as in the middle plane of the Rhodotron. Thus, the dynamics of the electron beam in these planes will also replicate the dynamics of the electron beam in the Rhodotron if the synchronization conditions are met. Many articles regarding the Rhodotron researched the electrons' movements in this type of accelerator. However, some aspects of the Rhodotron's dynamics were not mentioned in these articles.

For instance, since Rhodotrons' RF generators have accordant power and in frequencies of 100-200 MHz, the accelerator cavities are sufficiently large and demand significant cooling. Their radiuses are up to 1-1.3 meters. The electrons crossing the first gap have unrelativistic speeds and participate in the grouping process. The large electrical gap, where the beam's electrons first accelerate after injection from a non-relativistic injector (80-100 kV), provides good electron capture into the acceleration region. Simulations of electron dynamics in this gap demonstrate that all of the electrons in an interval $\pi/2$ in length are grouped in an interval that is four times smaller, with a length of approximately $\sim\pi/8$, providing a 25% capture rate. This is shown in FIG. 1.

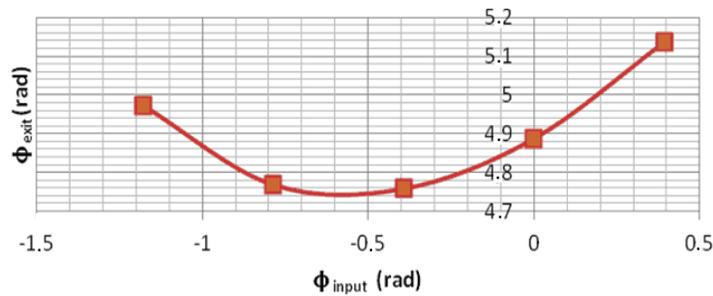

FIG. 1

Since each pass through the cavity provides approximately 1 MeV of energy, the electrons captured in the acceleration region will be relativistic and have velocities near the speed of light after the first pass. This assumption allows the integration of the radial equation of the electrons' motion in the cavity for the next pass in the finite view.

$$P_{(n+1)} = P_{(n)} + \Delta P_0 \cos(R\omega/c + \phi), \qquad (1),$$

where $P_{(n+1)}$ and $P_{(n)}$ are respectively the impulses of motion after n + 1 and n passes through the cavity and $\Delta P_0$ is the maximum of the change in the electrons' impulse in the cavity for one pass.*
Assuming that the electrons enter the cavity at the same phase during each pass, the total impulse of the motion of the electrons after all of the passes is

$$\Delta P_0 = e_0 \int E_\rho(\rho, t, \varphi_0) dt = -2(Ae_0/c)\cos(\varphi_0 + R\omega/c) \int (1/\varrho)\sin(\varrho) d\varrho$$

where c is the velocity of light and u is velocity of the electron, $e_0$ is the charge, $m_0$ is the mass of the electron, $p = m_0\gamma u$, $\gamma = (1-\beta^2)^{-1/2}$, $\beta = u/c$, $\omega = 2\pi f$, f is the resonant frequency, $\varphi_0$ is the input electron phase, $E_\rho(\rho, t, \varphi_0) = (1/\rho)$, $A\sin(\omega t + \varphi_0)$ is the radial component of the electric field in the cavity, and r and R are the radius of the inner and outer conductors of the coaxial cavity, respectively.

---

*

$$P_{(n)} = P_{(0)} + n\, \Delta P_0 \cos(R\omega/c + \phi_0 + \Delta\phi), \qquad (2),$$

where $R\omega/c + \phi_0 = 0$ is the phase of the maximum acceleration and $\Delta\phi = \pi/8$ is the boundary of the acceleration capture. A simulation of the initial acceleration stage for four passes through the cavity is illustrated in FIG. 2. The trajectories of the electrons with $\phi = -1.5708$ (Series 6) and $\phi = 0.7854$ (Series 7) are also shown. These trajectories (Series 6 and 7) exit the acceleration region.

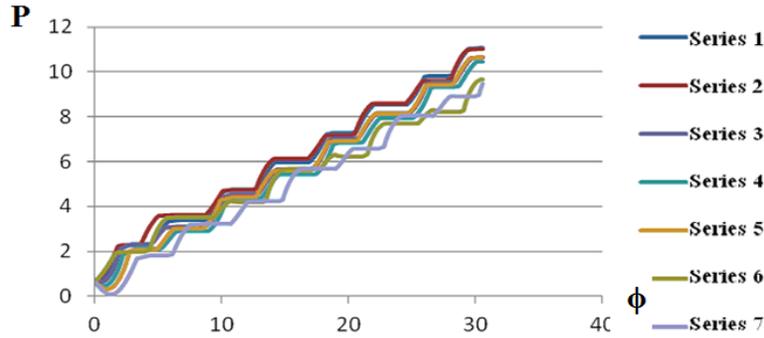

FIG. 2

FIG. 3 shows a simulation of the electrons' motion. There is a zone of longitudinal phase stability that is approximately $\pi/8$ or slightly higher

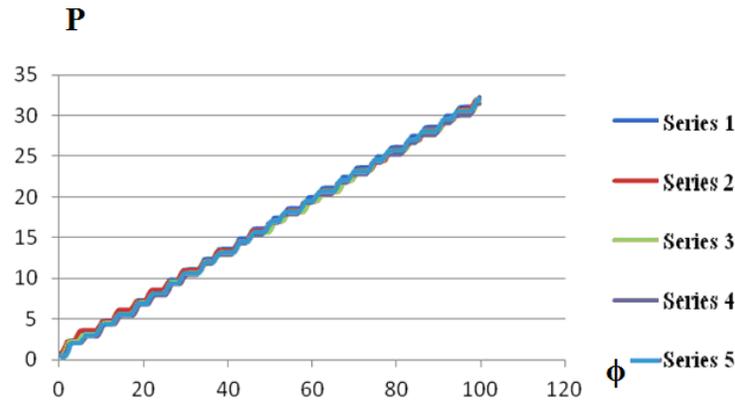

FIG.3

Without increasing the accelerator's frequency, the Rhodotron allows to obtain approximately 10-12 passes and the miniaturization of these types of accelerators further reduces this number. This restricts any increase in the electrons' energy and narrows the scope of the accelerator's use. This was why the next generation of Rhodotrons was designed [2] for some applications requiring an increase the electrons' energy at the accelerator exit. Design of electron accelerators with parameters of 40 MeV and more has been intended for change of nuclear reactors in manufacturing the medical isotope Mo99.

### Modern production problem of medical isotopes

Production of these isotopes was suspended because many nuclear reactors reached the end of their terms of operation without renewal prospects or the construction of new reactors. This problem was addressed in "Medical Isotope Production Without Highly Enriched Uranium" from the Committee on Medical Isotope Production Without Highly Enriched Uranium, National Research Council [3], published in 2009, which discussed different methods of solving this problem. Supplying small medical cyclotrons (up to 25 MeV and approximately 20 kW in proton beam) for medical institutions that need medical isotopes was implemented in Canada under the CycloMed99 program [4].

Several technologies are available that use electron accelerators to produce industrial scale isotope Mo99 to help solve this problem. These accelerators produce bremsstrahlung to irradiate different types of targets to manufacture isotopes. The power of the accelerator and the thermal resistance of the tungsten targets that convert the electron beam into bremsstrahlung are the main parameters that define the productivity of devices that produce isotopes.

The point of view onto the full needed power of the group of accelerators that are using in a center of manufacturing medical isotopes in global scale is univocal. If these devices have power of up to 20-40 kW for serving one medical institution, so the industrial center's own electron accelerators must have a total power of approximately 1-1.2 MW or the center must have a minimum of 40-50 accelerators (such as linacs). Many searches taking into account the world needs are confirming these numerals. For instance, the total power of the center's accelerators was confirmed by studies on designing new technology to produce isotopes (Mo99 and others), using an approximately 500 kW electron accelerator [5].
This designates necessary business expediency for the center's total accelerator power for isotope production. Electron accelerators such as linacs are less applicable for this task because they very rarely have power greater than 90-100 kW. Electron accelerators designed by IBA could be more perspective accelerators for this task, taking into account their power.

Belgian company IBA has produced a series of electron accelerators similar to Rhodotrons (TT-50, TT-100, TT-300, and TT-1000) with 50, 100, 300, and 1000 kW of power, respectively, but their electron beam exit energy is only approximately 10 MeV. The electrical efficiency of linacs is equal to 15-20%. The efficiency of Rhodotrons is significantly better because the electron beam crosses the same accelerating cavity several times, what is the significant advantage of these accelerators because their electron beam exit power is high (close to 600 kW) and the losses in the cavity walls are approximately 400 kW.

As previously noted, the Rhodotron's design does not allow more than 10-12 electron beam passes through the cavity. Therefore, the electron beam output energy level, 40-50 MeV, may be obtained only by increasing the electromagnetic field level in the cavity. It has been demonstrated in the accelerator's project (TT-300 HE) by the IBA [6]. IBA and NorthStar announced a joint venture to supply two TT 300

HE electron accelerators (40 MeV and 125 kW) in 2020 and six accelerators in 2024-2026 (for a total acceleration power of 1 MW).

The four-fold increase in the cavity's electromagnetic field in the TT 300 HE will increase the losses in the cavity wall by 16 times. If the cooling system is similar to the TT-1000 (the most powerful), it is necessary to change the CW mode to the impulse mode with the ratio of durations of the pulse and the pause as 1/16. Then the accelerator's 2400 kW output power will also decrease by 16 times or up to approximately 150 kW, coinciding with the parameters of the TT-300 HE accelerator, which has a cavity, based on the TT-1000 cavity.

The total losses in the walls of the eight cavities of TT-300 HE accelerators reach 3.2 MW and their total exit power reachs 1 MW. The electrical efficiency of these devices (including 50% efficiency of the eight RF generators) is approximately 12%, which is the main disadvantage of the NorthStar/IBA project because the cost of electricity going out to the losses will be $5.5 million per year. Estimating the total global market for medical isotopes as $300-350 million, the electricity losses are too high for one center for medical isotopes. These excessive annual electricity expenditures constitute approximately 10% of the cost of all eight accelerators (TT-300 HE), and in 10 years, the project cost will double even without other expenses. A more exact estimation of the possible NorthStar/IBA project losses of and calculations of the investment income can be obtained via the previously described computations [3].
Low-efficiency accelerators such as linacs or Rhodotrons (such as IBA's TT-300 HE) are the reason for search of new plans, but with the absence of other devices, even these projects are attractive for replacing nuclear reactors.

### The first proposal

Accelerators designed with coaxial cavities using higher-order TEM modes radically changes the situation. They have two or more planes where the radial electrical field in the cavity is the maximum [7] and the trajectories of the electron beam passes can be placed in these planes. The Multirhodotron is an example of such an accelerator.

In the Multirhodotron, the trajectories of the electron beam in the coaxial cavity are in several planes similar to FIG. 4 (a and b) and are placed where the electric field has the maximum amplitude.

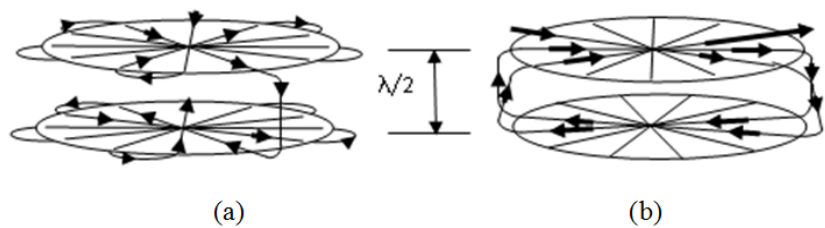

(a)           (b)           FIG. 4

Each of these designs has advantages and disadvantages. Electron bunches will cause an increase in the acceleration with periodicity T in FIG. 4(a), where T = 1/f and f is the frequency of the accelerator cavity. In the second variant, the beam's bunches will accelerate with a periodicity of 1.5 T, but in this case there are some trajectory intervals during transferring the electrons from one plane to another that can be used for the placement of additional focusing elements.

An accelerator based on FIG. 4(a) with two planes for electron acceleration was presented in [8], but this variant increased the number of passes through the cavity by only in two times and had an electron beam exit energy of only approximately 20 MeV.

The estimation of the topography of the electron trajectories in FIG. 4(a and b) assumes that variant (b) enables more passes crossing the cavity. In FIG. 4(a), similar to the Rhodotron, the trajectories in most parts are in the plane that is perpendicular to the accelerator's vertical axis, and the bending magnets are also in this plane. But the magnet sizes in this plane are defined by their radius of turn and electron energy. Therefore, the size of bending magnets in this plane are significantly larger than in the perpendicular plane.

The vertical size of the magnets may decrease up to 150 mm because the electron energy and the radius of turn will define the magnet size in the direction that almost coincides with the accelerator's vertical axis. This enables the placement of a large number of the bending magnets along the accelerator's outside circle in FIG. 4(b). When the radius of the cavity's outer cylinder is one meter and more, this number will reach 40-60 units. The electron trajectory in this type of accelerator is illustrated in FIG. 2, where the number of magnets is 22 (for simplification of picture).

Synchronization is necessary to maintain the electrons in the stable phase and they must enter the cavity at the same phase for each pass (taking into account that the electromagnetic field phases in the upper and bottom planes are opposite). This condition may be met by using the distance from the accelerator's axis to the placement of the magnet. The magnetic field in the bending magnets are acceptable for turning radiuses of approximately 150 mm and energies of 10, 20, 30, and 40 MeV. The magnet induction in the magnet gate is

$$B (T) = m_0 c^2 \gamma / cR = 0.011\gamma \qquad (3)$$

or approximately 2400, 4600, 6900, and 9200 G. These magnet induction rates enable designing the bending magnet as the permanent magnet, using (Sm-Co) or (Nb-Fe-B) fusion in accordance with accelerator's modern techniques where these magnets' electrical coils may regulate the field in the gates to the needed values.

Increasing the cavity up to λ will increase the losses in the cavity walls almost two-fold compared with the original Rhodotron but the Multirhodotron eliminates this problem. The Multirhodotron's acceleration voltage may be decreased but the number of passes through the cavity will be increased while maintaining its full acceleration. Decreasing the voltage by 25% will decrease the losses almost two-fold and will return to the former level at approximately 400 kW. Therefore, 60 passes through the cavity will provide an accelerator output of approximately 45 MeV with an electrical efficiency of approximately 60-70%. The approximate characteristics of the Multirhodotron for producing medical isotopes are shown in Table 1 and its structure in FIG. 5

Table 1.

| | | | | |
|---|---|---|---|---|
| 1. | Number of planes | 2 | 8. Breakdown voltage (kV/sm) | 37.5-40 |
| 2. | Output energy (MeV) | 45 | 9. Number of passages | 55-60 |
| 3. | Output electron current (A) | 0.015 | 10$^*$. Wall losses (kW) | 400 |
| 4$^*$. | Frequency (MHz) | 100 | 11. Energy per passage (MeV) | 0.75-0.8 |
| 5$^*$. | Power of RF generator (kW) | up to 1000 | 12. BBU threshold (A) | 0.017 |
| 6$^*$. | Mode of generator | CW | 13$^*$. Output power (kW) | 675 |
| 7$^*$. | Resonance cavity | coaxial cavity | 14. Overall dimensions (m) | 4.5 x 6 |

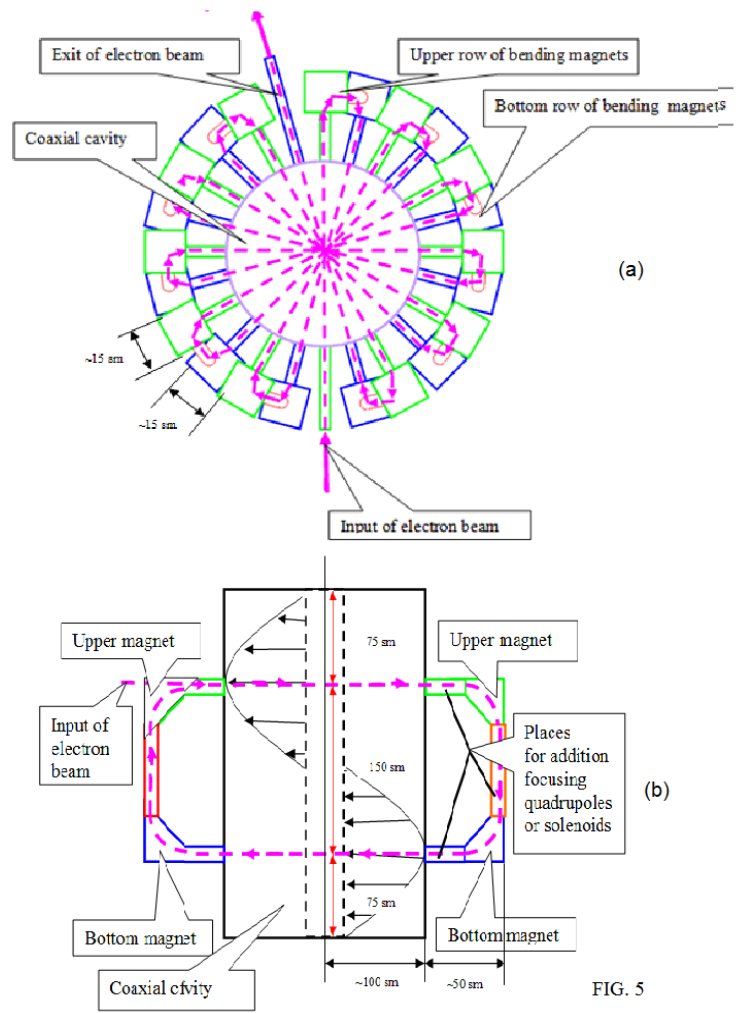

FIG. 5

In Table 1, the lines with asterisks (*) replicate the parameters of the TT-1000 accelerator. The majority of the accelerator's systems will be identical or very similar. Therefore, this variant will be a near replication of IBA's accelerator . The parameters ensure that the cost of the accelerator will be the same or slightly higher than that of the original Rhodotron TT-300 HE ($7 million). This will produce an industrial amount of Mo99 by one or two Multirhodotrons instead of eight accelerators as TT-300 HE.

If an annual cost of electricity of approximately $5 million is taken into account, the Multirhodotron's advantages will be obvious to any investor and the saved resources will be enough to redesign the Rhodotron TT-300 HE (TT-1000) into a Multirhodotron.

Another problem is beam breakup (BBU) instability, which can limit the total current of the electrons in the accelerator's beam. In the Rhodotron (TT-1000), this current is approximately 100 ma. The current limit depends on the number of electrons in the accelerating beam and the number of the beam passes through the cavity. Therefore, a six-fold increase in the number of beam passes can be adjusted by a six-fold decrease in the beam current. This can also be achieved by redesigning the accelerator.

The Multirhodotron's high beam output power demands a very accurate design of the converting tungsten targets to avoid damaging them. Historically, the Rhodotron's electron accelerator was designed as a very powerful bremsstrahlung source for irradiating different subject placed on the standard pallet surface, enabling the irradiation of objects 100 x 120 x 200 sm. The most powerful Rhodotron, the TT 1000 with an exit power of approximately 600 kW, can work very accurately and carefully with the tungsten targets in this case so Multirhodotron will be able to carry out it of course too.

The positive features of Rhodotron and Multirhodotron can be united for approaching to new horizons of accelerator's projects and application technologies. Simultaneous increasing the number of passes through the cavity like in Multirhodotron jointly with the increasing of the acceleration voltage in cavity allows assuming that the project of electron accelerator with new combination of parameters might be implemented. The practice of company IBA in the project of Second Generation Rhodotron showed that the four-fold increasing of the acceleration voltage in the coaxial cavity does not give the arcs and breakdowns on the surface of inner cylinder of cavity. The Table 2 with parameters of such electron accelerator on the base of Multirhodotron can be as a result of above thoughts. The result considered below in Table 2 relates to the five-fold increase of acceleration voltage in comparison to Table 1.Certainly, the transition in working of accelerator to the impulse mode will give the reduced an efficiency of this device like in the project TT-300 HE.

Some physical effects can limit this transformation of parameters. It is connected to synchrotron radiation of relativistic electrons of high energy on moving along of curvilinear orbits.

Table 2.

| | | | |
|---|---|---|---|
| 1. Number of planes | 2 | 8. Breakdown voltage (kV/sm) | 200 |
| 2. Output energy (MeV) | 225 | 9. Number of passages | 55-60 |
| 3. Output electron current (mA) | 0.6 | 10. Wall losses (kW) | 400 |
| 4. Frequency (MHz) | 100 | 11. Energy per passage (MeV) | 3.75 |
| 5. Power of RF generator (kW) | up to 700 | 12. BBU threshold (A) | 0.017 |
| 6. Mode of generator | pulse mode | 13. Output power (kW) | 175 |
| 7. Resonance cavity | coaxial cavity | 14. Overall dimensions (m) | 4.5 x 6 |

**The second proposal**

Revisiting the electron dynamics in the Rhodotron's accelerating gap, the electrons accelerate at the first interval by entering the external cavity cylinder to entering the inner cylinder and in the second interval by exiting the inner cylinder to exiting the external cylinder. This occurs because the electromagnetic field phase changes to the opposite phase when the electrons move inside the inner cylinder. Thus, the electric and magnetic fields are forming the focusing forces during the first interval and are forming the defocusing forces during the second interval, so the cavity, as a focusing element, forms a doublet (FOD) of the focusing and defocusing wide lenses. In any lens sequences (FOD or DOF), these lenses will cause the focusing effect. In this case, the electromagnetic field also focuses the electron beam when the field decelerates the electron beam while passing through the cavity.

Besids it, the simulation of the electron dynamics shown in FIG. 6 demonstrates that an area of phase stability is present during the deceleration of the electron beam in the coaxial cavity, similar to the region of phase stability in electron beam acceleration process in the Rhodotron and Multirhodotron.

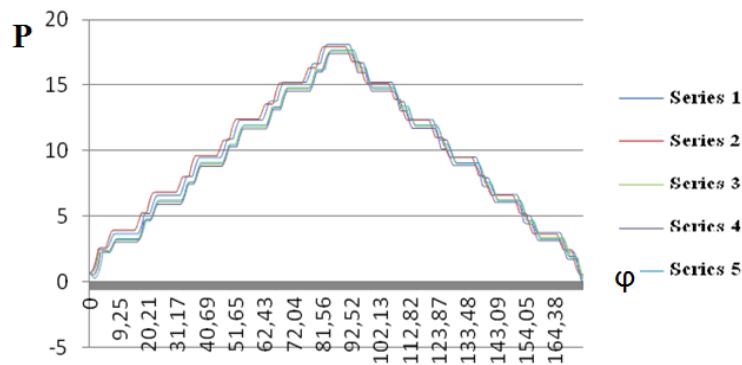

FIG. 6

This type of accelerator can transform the accelerated electron beam energy back into the electromagnetic field energy for recuperation, analogous to the energy recovery process in the linac and the JAERI ERL project [9]. The decrease in the electron's energy in this beam can continue almost to zero.

In the Multirhodotron, the accelerating part of the beam and the decelerating part of beam must not cross the cavity in the same place, as in the linac, because the beam's acceleration is provided in one group of cavity gaps and the beam's deceleration is provided in another group of gaps in the same cavity. This distinguishes the Multirhodotron from all other types of accelerators implementing ERL techology.

The Rhodotron is the most powerful electron accelerator with the exception of accelerators with superconductive cavities. For instance, in the JAERI ERL project, the accelerator with such cavities has a 20 MeV electron beam and a 40 ma current, that is, a 800 kW beam. Usually, energizing the FEL wiggler provides a light flow power approximately 1-2% of the beam power. A 12 kW infrared diapason (10 microns) has been obtained experimentally but further increasing the light flow is difficult as it requires an increase in the electron beam power at the wiggler input. This diapason of wave-length is very attractive because it coincides with the light flow wavelength of a $CO_2$ laser, and most lenses and mirrors are designed with a 50-100 kw of light flow power.

This Multirhodotron feature allows to revise the idea of using an electron accelerator to energize powerful FEL, as the Rhodotron used in [10]. Both the Rhodotron and Multirhodotron may be used for this purpose. The 20 MeV beam in the accelerator (TT-300 HE) enables to use one half of the electron beam such as acceleration passes and the second half of the passes recuperates the electron beam energy after the wiggler.

This variant produces an average light flow of approximately 1 kW and 16 kW in impulse and the last fact requires in an impulse of 3-4 diacrodes (TH-628). Therefore, this variant's total efficiency is very small for practical purposes.

Another possible variant of the accelerator without the recuperation of the electron energy might be TT-1000 but with a two-fold increase in the cavity field in the CW mode that provides 12 kW light flow level and 1600 kW in cavity's loss. But this requires energizing three diacrodes to provide 2800 kW in the cavity. The total loss in cavity and on three anods of diacrods is 4000 kW (1600+ 3*800) and the total efficiency is approximately 0.23%. This is also very small. Therefore, neither the impulse mode nor the increase in the cavity field provide practical results for using an accelerator with a coaxial cavity such as the Rhodotron to energize FEL.

Increasing the number of electron beam passes through the coaxial cavity significantly changes the situation. If the Multirhodotron parameters in Table 1 are used, then the number of passes crossing the cavity may be divided into three groups. The first consists of 27 passes, the second consists of 27 passes, and the third consists of six passes. The third group will be described further.

The first group has a 20 MeV accelerated beam to move the beam out of the accelerator into the wiggler. The beam power is 340 kW and the second group decelerates the beam after the wiggler to almost zero. In this case, the wiggler obtains approximately 4.5 kW of the light flow and the efficiency is 0.375%. However, the Multirhodotron has a special resource (increase of accelerator's current) that might be used in TT-300 HE and TT-1000 devices, but in a lesser degree because they demand excessive power from RF generators.

Both the Multirhodotron and Rhodotron have the limited beam currents because BBU instability appears in the accelerator if the current exceeds this limit. Despite this the focusing forces in the coaxial cavity restrain the beam's radius if even its current exceeds 10 and more amperes in the fields that provide approximately 1 MeV per pass in the cavity [11].

Theoretically and experimentally, the mechanism of BBU instability can be obtained as the energizing of the higher-order modes (HOMs) than the working oscillations in the coaxial cavity. Some of these modes have the maximums of magnetic field in those planes where the radial electric field of the working oscillation causes the acceleration of the electron beam. The magnetic field action in these modes translocate electrons from these planes and transfers them to areas where the electric field of the HOMs differs from zero, where the electron beam will be decelerated by these HOMs, increasing the level of fields of these modes.

There are simple methods of fighting this energizing mechanism that are often used to manage the Foucault currents induced in the conductive surfaces of the cavity walls. In this method, narrow slits are been made on the surface where the Foucault currents are at a maximum and these higher-order oscillations will not be energized.

In this case, two or more narrow slits must be made in the coaxial cavity's inner and external cylinder where the working mode has the maximum radial electric field, that is, where the electrons accelerate. This allows increasing the accelerating beam current up to 2-3 amperes while maintaining the acceleration. The working oscillation will not be changed by the slits in the cavity because the working mode magnetic field has null value and the slits do not hinder the working oscillation energization.

The Multirhodotron's accelerated beam with a current of approximately 2 amperes has 40 MW at the accelerator exit and so provides 400 kW of the light flow at the wiggler without any changes in the circuit of the Multirhodotron's RF generator. Inhibition of BBU instability in the Rhodotron and Multirhodotron more than 10 times increases their output power.

Indeed, the FEL's electrical efficiency is defined by:
$$\eta = \eta_L W_L/(W_C + \eta_L W_L), \qquad (4)$$

where $W_C$ is the average power of the losses in the accelerator and wiggler, $\eta_L$ is the efficiency of the electron radiation in the wiggler, and $W_L$ is the average power of the electron flow.

The FEL in this case will have an efficiency of approximately 25% if the losses in the wiggler are ignored. The electron accelerator's efficiency will be almost 100%, so $\eta = 40/41.2 = 97\%$. However, it is possible to obtain still big efficiency for based on the Multirhodotron FEL, if the addend in denominator in (4) will be significantly more $W_C$. How to reach these conditions by means of using the feature of Multirhodotron is our second proposal.

The wavelength of the light in the wiggler is defined by:

$$\lambda = (1 + \alpha^2_w/2) \lambda_0/2\gamma^2 \qquad (5)$$

where $\alpha_w = e_0 B_0 \lambda_0/(2\pi m_0 c)$ and $\lambda_0$ is the period of the undulator.

According to formula (5) the electrons in the beam will have not large energy differences after the sequential crossing of the cavity several times therefore if directing the electrons after each cavity pass into 6-7 wigglers alternately, from 28th up to the 33rd pass. The total power will be 2.4-2.8 MW that is significantly more than Wc, but with different wave-length in each wiggler. During this process, each cavity pass restores the electrons' energy after the loss of energy in the previous wiggler. Concomitantly, the conterminous meanings of the light wavelengths can be obtained, changing the magnetic fields in all of the wigglers according to formula (5).

In the next group of passes through the cavity, beginning from 34th, the energy of the electrons is transformed back into the energy of the electromagnetic field in the cavity. But the losses of electron energy by means of the light radiating are needed in compensation from the powerful RF generator

The total electrical efficiency of the FEL in accordance with formula (4) is equal to 70% approximately and this efficiency makes the device very attractive in practice and business. A schematic of the FEL structure based on the Multirhodotron is shown in FIG. 7(a and b).

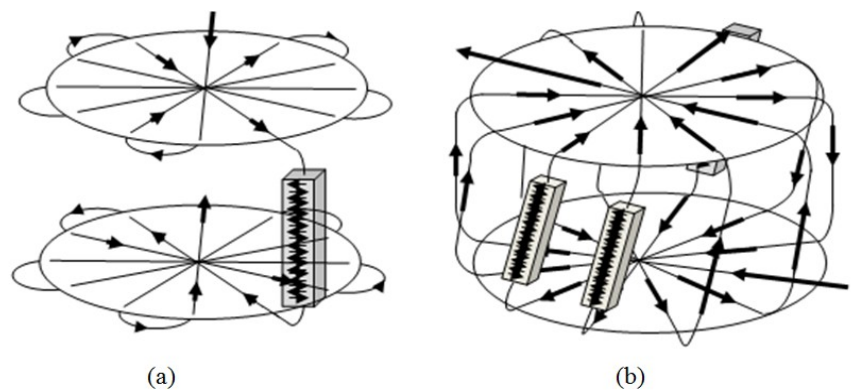

(a)     (b)     FIG. 7

# The third proposal

Another important application of the Multirhodotron with the return of the electron energy is the use of the electron beam for "electron cooling" technology. Most ion and proton accelerators and colliders are equipped with devices for "electron cooling" to increase the luminosity of the ion and proton beams.

In the 1970s, electron cooling systems used 2-4 MeV DC accelerators with currents up to 1 ampere. In 2000, new ideas were suggested for electron cooling technology. This technology uses a bunched electron beam with energy of approximately 54 MeV and an average current up to 0.5 amperes that is produced by the superconducting linac using ERL technology. The accelerator has four sections with five cells in each for a frequency of approximately 705 MHz, as described in [12], for the planned cooling of a beam at the Relativistic Heavy Ion Collider at Brookhaven National Laboratory (BNL).

The parameters of the electron accelerator are in the range that is covered by the Multirhodotron. The recuperation technologies of the beam's energy in projects such as JAERI ERL also utilize superconductivity linacs but their implementation is complex.

At 705 MHz, the length of the bunch is approximately 20 mm and the bunch must be debunched up to 60-80 mm before the cooling proton or ion beam is used. After cooling, the same bunch must be bunched again up to 20 mm before the accelerator recuperates the electron energy. For the Multirhodotron and the 200 MHz generator, length of the bunch must be 75 mm and the constant bunching-debunching process is unnecessary.

The cooling process requires $(1-3)10^{10}$ electrons in the bunch for BNL's project or the charge of the bunch is 1.6-4.8 nanocoulombs according to the average current of beam (0.32-0.96) A in the accelerator. These requirements lay in the possible diapasons for the Multirhodotron if to overcome the BBU instability in the accelerator's coaxial cavity, but as previously mentioned, there are simple and effective methods to prevent BBU instability in the coaxial cavity. Therefore, the application of the Multirhodotron for "electron cooling" can be successfully accomplished without using a superconductive linac.

The large finite energy of the electrons in this case will require an increase in the number of passes through the cavity to 80-110 (or 1-1.35 MeV per pass), where the first half of the passes accelerate the beam and the second half decelerate the electrons. The Multirhodotron's electron cooling may be technically implemented, if the RF generator's frequency increases up to 200 MHz and a cavity length up to $2\lambda$, to act upon the electron beam in four planes.

The total sizes of the cavity and accelerator will be somewhat miniaturized in this case and also will be decreased the number of possible crosses of the cavity in each plane up to 28-30 passes. The losses in the cavity will also increase significantly despite the reduction in the square of the surface of the cavity's walls.

The losses in the cavity walls in the analyzed case are not small because of the skin effect and the two-fold wavelength of the accelerator's coaxial cavity. The accelerating field level (1.35 MeV per pass) will cause nearly two-fold losses too. Therefore, the RF generator in this case may create problems with radio-tubes at an output power level of approximately 2-3 MW. This level of power for energizing the cavity requires very powerful tubes or the use of several tubes in parallel as described in IBA's patent EP 2509399 (10-10-2012) [13].

However, the Multirhodotron resolves this problem using technology known as TBA. The electron energy of one beam is used for the electron acceleration of the second beam. In the case analyzed herein, our patent CA 2832816 (11-12-2013) can be implemented if 2-3 additional passes through the cavity are used [14]. This method does not require additional RF tubes to increase the power entering the cavity.

The principle of action of the generator is that the high-voltage DC relativistic injector injects continuous electron beam into the first pass of the cavity, energized by an auxiliary generator. This injector's electrons have initial energy equal to the increase in the energy that the electrons must obtain in each pass through the cavity in nominal operation. It may be 1.0-1.5 MeV or close. After crossing the cavity, the beam will not change uniform continuous character, but half of the electrons will have energy greater than the injection energy. The second half will have energy less than this level of energy. The changes in the impulse of motion of the electrons will have sine wave characteristics and will depend on the input phases of the electrons in the beam.

$$P = P_0 + \Delta p = P_0 + \Delta p_m \sin(\varphi_0 + \chi), \qquad (6)$$

where $P_0$ is the initial impulse of the motion of the electrons injected from the high-voltage relativistic injector, $\Delta p_m$ is the amplitude of change of the impulse of motion of the electrons in the cavity beam, $\varphi_0$ is the input phase of the electrons, and $\chi$ is the transient factor of the electrons in the cavity.

The impulse P of each electron is interlinked with the total energy E of the electron by means of the relativistic invariant.

$$P^2 - E^2/c^2 = -m_0^2 c^2, \qquad (7)$$

where $m_0$ is the electron rest mass, c is the velocity of light, and $E = T + m_0 c^2$, where T is the kinetic energy of the electron.

Formula (7) calculates the kinetic energy of the electron if the impulse of the electron is known and hence the integral of the kinetic energy along the phase φ = (φ₀ + χ) from 0 to 2π (or along the length of the beam) may be calculated analogously with the integral of the impulse. If $\Delta p_m/P_0 \ll 1$, the increment of the integral of the kinetic energy of the beam's electrons for $0 < \varphi < \pi$ is approximately equal to the decrease in the same integral for $\pi < \varphi < 2\pi$, so the beam does not load the cavity and hence the accelerating part of the beam, that corresponds to phases $0 < \varphi < \pi$, is accelerated by means of losses of kinetic energy of the electrons in the other part of the beam that correspond to phases $\pi < \varphi < 2\pi$. The resultant distribution of the kinetic energy $T_1(\varphi)/T_0$ and $T_2(\varphi)/T_0$ for both parts of the beam accordingly $0 < \varphi < \pi$ and $\pi < \varphi < 2\pi$ are shown in FIG. 8 for $\Delta p_m/P_0 = 0.1$, 0.5, and 1 and $P_0^2 - (T_0 + m_0c^2)^2/c^2 = -m_0^2c^2$.

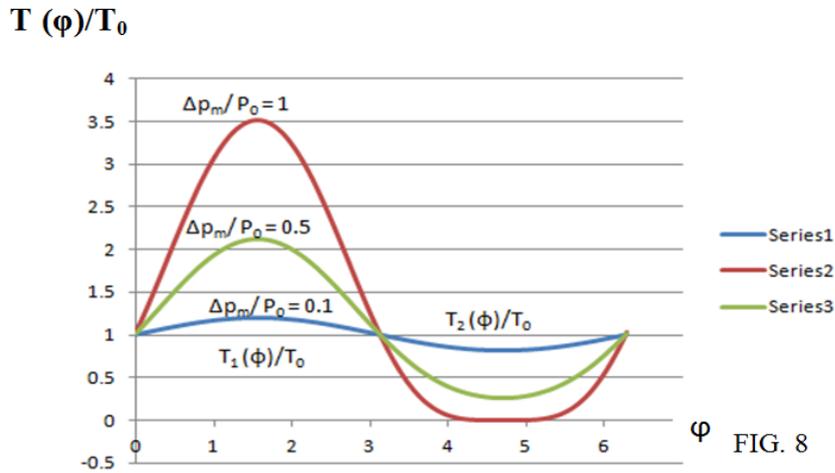

FIG. 8

$W_1$, the integral of kinetic energy $T_1(\varphi)/T_0$, will be significantly greater than $W_2$, the integral of $T_2(\varphi)/T_0$ when $\Delta p_m/P_0 = 1$. The relationships between the increasing and decreasing integrals of the kinetic energy of the beam's electrons are shown in greater detail in FIG. 9 as a function of $0 < \Delta p_m/P_0 < 1$. The results are given in FIG. 9 in relative forms, assuming that $W_0$ is equal to the integral of the kinetic energy of the initial beam for the interval $0 \leq \varphi \leq 2\pi$ before crossing the cavity.

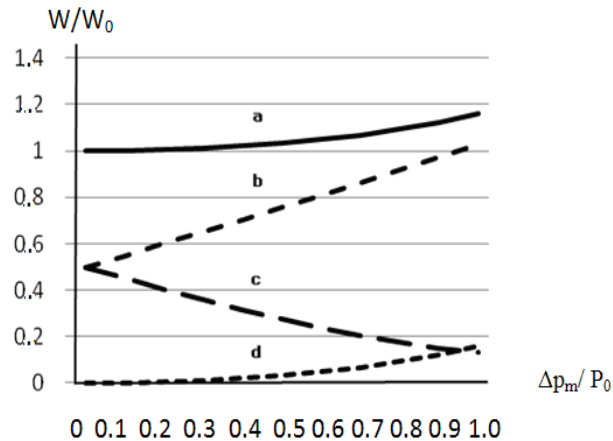

FIG. 9

The integral of the beam's kinetic energy along the length of the beam for the interval $0 \leq \varphi \leq 2\pi$ ($W = W_1 + W_2$) after crossing of the cavity is not equal to the same integral of the kinetic energy when the beam has not yet crossed the accelerated cavity. This is indicated by line a. All of the integrals in FIG. 9 have been calculated using numerical integration. In FIG. 9, for $\Delta p_m/P_0 = 1$, the accelerated part of the continuous beam (line b) concentrates almost all of the kinetic energy of the initial beam. The rest of the beam for $\pi \leq \varphi \leq 2\pi$ (line c) might be deleted from the continuous beam without essential losses of the beam's initial input power. Line d shows the power required from the external generator.

The efficiency of the conversion of electric energy to kinetic energy of electrons is calculated as

$$\eta = W_1/(W_1 + W_2) = 1/(1 + W_2/W_1),$$

where $W_1$ and $W_2$ depend on $P_0$ and $\alpha = \Delta p_m/P_0$.

For any $P_0$ value, the integral $W_2$ is the minimum at $\alpha = 4/\pi$. In this case, the efficiency $\eta$ is 90% when the initial energy of the beam's electrons is 1.0-1.5 MeV. When $\alpha = 4/\pi$, the reverse conditions of the electrons' velocity are met for the decelerated electrons and the beam of returning electrons appears. This may require additional technical design for their neutralization and therefore it is better to use $\alpha = 1$ to avoid these complications. Under this condition, the efficiency $\eta$ is 87-88% for the same initial electron energies and differs insignificantly from the previously described minimum case. If the decelerated electrons are extracted out of the beam by devices such as a mass spectrometer and moved into a dump, then the rest of the electrons in the beam that accumulate almost the total energy of the initial injected beam can be used to energize the cavity like the modulated relativistic electron beam maintaining significant power inside.

In the Multirhodotron, when the injected relativistic electron beam crosses the cavity the first time, the energy of the electrons changes. After the beam transforms into a modulated relativistic electron beam via the magnet system with the transverse magnetic field, then the beam crosses the cavity once more, powering the electromagnetic field in the Multirhodotron's cavity.

A schematic of the additional generator based on the Multirhodotron is shown in FIG. 10. The accelerating (second) electron beam is not pictured. The first continuous electron beam with a current of approximately 1-2 amperes moves from the injector (4) to the accelerating tube (8). The beam (9) crosses the cavity (1) the first time. The beam then traverses the magnet system, leaving low-energy electrons in the dump (13) and transforming into a modulated electron beam. The beam then crosses the cavity a second time (18) and a third time (22) because when $\alpha = 1$, it is impossible to remove the total energy from the electrons during one pass. At the end, the electrons follow in the second dump (23). Between the second and the third passes, the beam (21) traverses the bending magnet (19) and is synchronized.

Thus, the Multirhodotron supplies the needed RF power because several additional generators can be conducted to the single cavity.

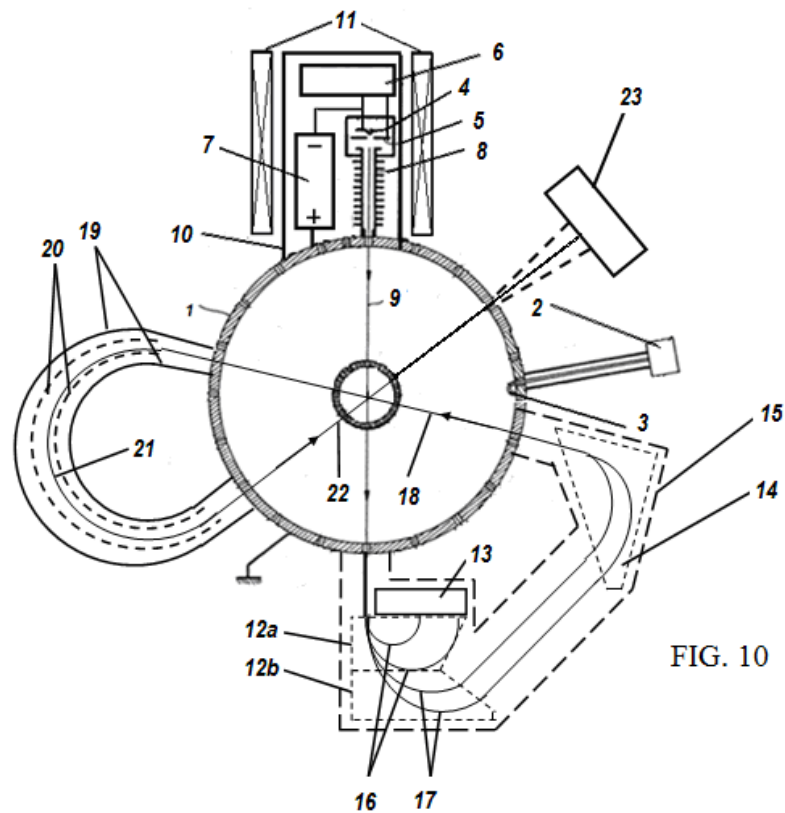

FIG. 10

**Conclusion**

The variations in the electron accelerator based on the coaxial cavity and the Multirhodotron's methodology provide new application prospects that include:
- electron accelerators with one cavity producing high-energy electron beams (approximately 30-60 MeV) in CW mode and up to 200 Mev in impulse mode
- high energy electron accelerators using ERL technology for accelerators without of superconductive cavities,
- electron accelerators not requiring of using radio-tubes (more than 1-2 MW) and energizing accelerator cavities with high efficiency in TWA technology.

These technical decisions were suggested for manufacture of medical isotopes in global scale, the designing of powerful FEL in IR diapason and device for "electron cooling". Also three themes were stated about researching new limits of energy and power of electron accelerator and radio-frequency generator and the methodology of struggle with BBU instability in accelerators with coaxial cavity.

**Acknowledgements**

We would like to greatly thank Dr. C. Ross for his helpful comments on the manuscript and the supporting in the process of demonstration of advanced features Multirhodotron, but also for advice connected to the necessary adequate assessing of features of usual linacs.

**References**

1. Patent US 5107221  ( 4/1992 )   N'Guyen et al.
2. IBA_Presentation_Aries_Annual_Meeting.pdf
3. Bookshelf_NBK215149.pdf
4. CycloMed99   https://www.triumf.ca/cyclomed99
5. Patent   US9721691    (Aug. 1, 2017)   Alexander Tsechanski
6. 290319_iba_industrial-northstar-en.pdf
7. Patent CA 2787794   (27-08-2012)   Gavich M., Gavich V.
8. 1-s2.0-S0168900216303588-main.pdf
9. R. Hajima et al. "JAERI ERL-FEL: status and future plans", The 14th Symposium on Accelerator Science and Technology, Tsukuba, Japan, November 2003.
10. J.M. Bassaler, C. Etievant  "Nuclear Instruments and Methods in Physics Research" A304 (1991) 177-180
11. J.M.Bassaler, http://www.iaea.org/inis/collection/NCLCollectionStore/_Public/25/003/25003090.pdf
12. Electron Cooling for RHIC Design Report, V.V. Parhomchuk, I. Ben-Zvi, Principal Investigators, 2001.
13. Patent EP 2509399 (10-10-2012)   M. Abs et al.
14. Patent CA 2832816 (11-12-2013)   Gavich M., Gavich V.